\documentclass[12pt]{article}
\usepackage{amsfonts,graphicx,amssymb,amsmath,amsthm,epsfig}
\usepackage{float}
\allowdisplaybreaks
\begin{document}
\title{\bf Generalized Ghost Dark Energy in $f(Q)$ Gravity}
\author{M. Sharif \thanks {msharif.math@pu.edu.pk}~and
\ Madiha Ajmal \thanks {madihaajmal222@gmail.com} \\
Department of Mathematics and Statistics, The University of Lahore,\\
1-KM Defence Road Lahore-54000, Pakistan.}

\date{}
\maketitle
\begin{abstract}
In this manuscript, we use the correspondence principle to construct
the generalized ghost dark energy $f(Q)$ model, where $Q$ is the
non-metricity. For this purpose, we use the
Friedman-Robertson-Walker universe model with power-law scale
factor. The energy density in this model linearly depends on the
Hubble parameter, with a sub-leading term of $H^{2}$, resulting in
an energy density $\rho_D=\xi H+\zeta H^{2}$, where $\xi$ and
$\zeta$ are arbitrary constants. We reconstruct interacting fluid
scenario of this dark energy with dark matter. The reformulated
$f(Q)$ gravity model shows how different epochs of the cosmic
history can be reproduced by this modified gravity model. The
physical characteristics of the model are discussed through the
equation of state parameter with the planes of
$\omega_{D}-\omega'_{D}$ and $r-s$. We also explore the stability
criteria for the interacting model using the squared speed of sound.
The equation of state parameter indicates phantom phase of the
universe and the squared speed of sound represents stability of the
interacting model. The $(\omega_{D}-\omega^{\prime}_{D})$-plane
indicates freezing region, while the $(r-s)$-plane corresponds to
the Chaplygin gas. We conclude that our findings coincide with the
latest observational data.
\end{abstract}
\textbf{Keywords}: $f(Q)$ gravity, Generalized ghost dark energy,
Cosmological evolution.\\
\textbf{PACS}: 04.50.Kd; 95.36.+x; 64.30.+t.

\section{Introduction}

General relativity (GR) is a theory of gravitation proposed by
Albert Einstein and is the most significant discovery of the 20th
century. This theory stands as Einstein remarkable scientific
legacy, providing a successful framework to unravel the mysteries of
our universe evolution and its concealed truths. Recent observations
indicate strong evidence about the universe accelerated expansion
phase \cite{1}. These observations provide that $95-96\%$ of the
universe contains dark energy (DE) and dark matter (DM) but only
$4-5\%$ of baryonic matter \cite{2}. The expanding universe is said
to be the result of exotic force with a large negative pressure,
known as DE. The cosmological constant in GR explains the expanding
universe but it has two fundamental problems like coincidence and
fine tuning. There are two main suggestions to investigate the
expanding nature, dynamical DE models and modified theories(MT).

The most fundamental variation of GR is the $f(R)$ theory, which is
derived from the Einstein-Hilbert(EH) action, by replacing the Ricci
scalar $(R)$ with an arbitrary function \cite{3}. Nojiri and
Odintsov \cite{4} established modified Gauss-Bonnet theory, termed
as $f(G)$ gravity, by introducing the generic function $f(G)$ into
the EH action, where $G$ is the Gauss-Bonnet invariant. In
particular, realistic models of $f(G)$ gravity have been formulated
for the observed cosmic accelerated expansion in the late-time
universe. The same authors \cite{2a} examined various types of
modified gravitational theories, including $f(R),~f(G)$ and $f(R,G)$
gravity theories, which are considered alternative theories for DE.
They revealed that certain theories within this category could
successfully pass the solar system tests and exhibit a robust
cosmological structure. The $f(R,\mathbb{T})$ theory of gravity
represents another extension of GR, in which $\mathbb{T}$ represents
trace of the energy-momentum tensor (EMT) \cite{4a}. Katirci and
Kavuk \cite{4b} made modifications to the $f(R,\mathbb{T})$ theory
by incorporating a non-linear term ($\mathbb{T}^{2} =
\mathbb{T}_{ab}\mathbb{T}^{ab}$) within the functional action,
leading to what is termed as $f(R,\mathbb{T}^{2})$ theory. Sharif
and Ikram \cite{4c} proposed the inclusion of a generalized
$f(G,\mathbb{T})$ function into the EH action to establish
\(f(G,\mathbb{T})\) gravity. They examined energy conditions within
the framework of the  FRW universe.

The general theory of relativity is a geometric theory associated
with Riemann space. There is another approach to obtain generalized
theories of gravity with a more general geometric structure which
can describe the gravitational field. Weyl \cite{5} developed a more
general geometry than Riemannian by introducing an additional
connection (length connection), which does not provide information
regarding the direction of the vector during parallel transport. In
this theory, the metric tensor's covariant divergence is non-zero,
which yields a novel geometric quantity is non-metricity. Dirac
\cite{6} generalized Weyl theory on the basis of two metrics but it
was ignored by physicists. Cartan \cite{7} made a significant
advancement in geometric theories, giving rise to a new generalized
geometric theory is Einstein-Cartan theory \cite{8}. The Weyl
geometry can be modified by incorporating torsion, dubbed as
Weyl-Cartan geometry \cite{8a}. Weitzenb$\ddot{o}$ck \cite{9}
suggested another mathematical development (Weitzenb$\ddot{o}$ck
spaces), whose characteristics are $\nabla_\mu
g_{\nu\lambda}=0,~T^\mu_{\nu\lambda}\neq0$, and
$R^\mu_{\nu\lambda\sigma}=0$, where the metric, torsion and
curvature tensors are represented by $g_{\nu\lambda}$,
$T^\mu_{\nu\lambda}$ and $R^\mu_{\nu\lambda\sigma}$, respectively.
The Weitzenb$\ddot{o}$ck manifold transforms into a Euclidean
manifold when $T^\mu_{\nu\lambda}=0$. These geometries possess the
crucial characteristic of distant parallelism, referred to as
absolute parallelism, or tele parallelism.

In the teleparallel theory of gravity, the metric tensor is replaced
with a collection of tetrad vectors. Consequently, the torsion is
used to characterize gravitational effects instead of the curvature.
This gives the teleparallel equivalent to GR, named as $f(T)$
theory, where $T$ denotes the torsion scalar \cite{10}. Haghani et
al. \cite{11} expanded the teleparallel gravity theory into what is
known as Weyl-Cartan-Weitzenb$\ddot{o}$ck gravity. The same authors
\cite{11a} proposed another extension of the
Weyl-Cartan-Weitzenb$\ddot{o}$ck and teleparallel gravity by
incorporating the condition of vanishing curvature and torsion in a
Weyl-Cartan geometry. The field equations in the $f(T)$ gravity
theory are of second order, compared to the $f(R)$ gravity theory,
which is a fourth order theory when examined through a metric
approach. The second-order nature of $f(T)$ gravity theory has a
significant benefit. This discussion leads to the fact that GR can
be geometrically divided into two equivalent ways: the curvature
description (both the non-metricity and torsion vanish) and the
teleparallel description (both the non-metricity and curvature
vanish). A third equivalent description is when the non-metricity
$Q$ in the metric signifies the change in vector length during
parallel transport. This acts as the fundamental geometric variable
characterizing the gravitational interaction properties, leading to
symmetric teleparallel gravity (STG), which is also known as $f(Q)$
theory \cite{12}. There is a great body of literature available
\cite{14} to discuss different geometrical and physical aspects of
this theory.

The non-metricity is a mathematical concept that emerges in theories
involving non-Riemannian geometries, providing an alternative cosmic
model without DE. Researchers have drawn attention to explore
non-Riemannian geometry, specifically $f(Q)$ theory, for various
reasons such as its theoretical implications, compatibility with
observational data and its significance in cosmological contexts
\cite{1aaa}. Recent investigations into $f(Q)$ gravity have revealed
cosmic issues and observational constraints can be employed to
indicate deviations from the $\Lambda$CDM model \cite{2aaa}. The
non-metricity scalar has been employed to detect the effects of
microscopic systems in \cite{6aaa}. Barros et al. \cite{8aaa}
analyzed the cosmic characteristics through redshift space
distortion data in non-metricity gravity. For further details, we
refer the readers to \cite{11aaa}-\cite{16aaa}.

The use of instabilities in quantum field theory suggest that the
model may not be well-behaved at the quantum level. Unstable models
can lead to inconsistencies when quantum effects are considered.
Despite such issues with ghost models, several researchers are
interested to explore GDE models for some other well-known
motivations. Comparing the predictions of GDE models with standard
models allows scientists to understand the differences and potential
observable effects. In summary, the use of GDE models in $f(Q)$
theory is a topic of theoretical exploration, but their realistic
viability and compatibility with quantum field theory must be
scrutinized to ensure a meaningful contribution to our understanding
of cosmology.

The Veneziano GDE has been suggested \cite{5bb} which is
non-physical in the standard Minkowski spacetime but has significant
physical consequences in non-trivial topological or dynamical
spacetimes. It produces a small vacuum energy density in curved
spacetime proportional to $\Lambda^{3}_{QCD}H$ and $\Lambda_{QCD}$
is the quantum chromodynamics $(QCD)$ mass scale. Therefore, no
further changes, degrees of freedom, or parameters are needed. For
$\Lambda_{QCD}\sim 100 MeV$ and $H\sim 10^{-33}eV$, the measured DE
density can be accurately ordered by $\Lambda^{3}_{QCD}H$. This
remarkable coincidence suggests that the fine-tuning issues are
removed by this model \cite{6bb}.

To address the issue of DE, cosmologists have proposed various DE
models, including phantom, tachyon, quintessence and Chaplygin gas
in different eras. These evolutionary models trace its roots back to
the GDE within the Veneziano field of QCD theory. In recent years,
this dynamic model has attracted significant attention due to the
interesting characteristics of Veneziano GDE, which may be
responsible for the current accelerated expansion. Cai et al.
\cite{14b} explored the Veneziano GDE contribution to the vacuum
energy density, demonstrating that it is not completely dependent on
$H$. Instead, an additional sub-leading term of $H^{2}$ must be
included in the energy density expression, results in the
generalized GDE (GGDE) model, designed to examine the early stages
of expansion. In this theoretical framework, the Veneziano ghost
maintains compliance with fundamental principles attributes of
renormalizable quantum field theory, avoiding any violations of
these essential aspects.

Sheykhi and Movahed \cite{27} investigated the consequences of the
interacting GDE  model within the framework of GR. They analyzed the
universe expansion by applying constraints to the model parameters.
Setare et al. \cite{28} introduced the interacting DE model within
the context of Horava-Lifshitz cosmology. Khodam et al. \cite{4aa}
examined the reconstruction of $f(R,T)$ gravity within the context
of the GGDE model. They also investigated the cosmic evolution by
analyzing various cosmological parameters. Ebrahimi et al.
\cite{5aa} examined the interacting GGDE in a non-flat universe,
discovering that the EoS parameter resides in the phantom era of the
universe. Sharif and Jawad \cite{15c} analyzed the dynamics of
interacting pilgrim DE models using standard cosmological
parameters. Additionally, they assessed the stability of the DE
model within the context of the FRW universe model. Ebrahimi and
Sheykhi \cite{30} proposed the detection of instability signals from
generalized quantum chromodynamics GDE. Pasqua et al. \cite{38a}
studied DE at a higher order of $H$ in the context of
$f(R,\mathbb{T})$ gravity. Their findings revealed that the model
demonstrates stability in the early universe but becomes unstable in
the current era.

Sharif and Zubair \cite{3a} applied the statefinder diagnostic to
both the phantom and quintom DE models. Jawad and Rani \cite{3b}
investigated the reconstruction of generalized ghost pilgrim DE in
the context of $f(R)$ gravity. Sharif and Nawazish \cite{38}
investigated the interacting and non-interacting DE models in $f(R)$
gravity. Sharif and Saba \cite{38b} studied the cosmography of GGDE
in $f(G)$ and $f(G,\mathbb{T})$ gravity theories. Lu, Zhao and Chee
\cite{20a} studied the cosmic characteristics in symmetric
teleparallel gravity (STG), revealing that the accelerating
expansion is an inherent property of the universe's geometry. Lazkoz
et al. \cite{20b} investigated observational constraints of $f(Q)$
gravity. Mandal et al. \cite{15} analyzed limitations on the Hubble
constant and the cosmographic functions in this theory. Shekh
\cite{15a} studied the dynamical investigation of holographic DE
models in the context of this theory. Mandal et al. \cite{8aa}
calculated the value of equation of state (EoS) parameter for
Pantheon samples as well as Hubble parameter. The result of $f(Q)$
model shows quintessential behavior that diverges from the standard
cold DM. Lymperis \cite{17} investigated the cosmological
implications through the effective DE sector in the same theory.
Solanki et al. \cite{18} discovered that the geometric extension of
GR could serve as a viable candidate for explaining the origin of
DE. Recently, the parametrization of the effective EoS parameter in
this framework has been explored in \cite{19}.

This paper reconstructs the GGDE $f(Q)$ model in the interacting
case using the correspondence scheme. We investigate evolution of
the universe using the EoS parameter, the squared speed of sound
($\nu_{s}^{2}$) and phase planes. We organize the article in the
following pattern. Section \textbf{2} presents the description of
this MT of gravity. In section \textbf{3}, we discuss the effects of
interacting DE and CDM in terms of red-shift parameter and apply a
correspondence technique between GGDE and $f(Q)$ gravity to
reconstruct GGDE $f(Q)$ model. Section \textbf{4} is dedicated to
explore the evolution of this model through cosmographic analysis.
Section \textbf{5} provides the summary of our results.

\section{Brief Sketch of $f(Q)$ Theory}

This segment briefly reviews the geometrical structure of the MT of
gravity on the basis of the spacetime. We use the variational
principle to formulate the field equations of the $f(Q)$ theory.

\subsection{Geometrical Foundation}

Weyl \cite{5} extended the Riemannian geometry (basis of GR) by
assuming that an arbitrary vector will change both its direction as
well as length during the parallel transport around a closed path.
The resultant geometric theory exhibits mathematical properties with
the vector $v^{\gamma}$ that are identical to those of
electromagnetic potentials. This implies that there might be a
shared geometric origin for both electromagnetic and gravitational
forces \cite{10}.

In a Weyl space, when a vector of length $y$ is parallel transported
along a very small path, say $\delta x^{\gamma}$, the change in its
length is given by $\delta y=yv_{\gamma}\delta x^{\gamma}$. However,
for a vector with a small closed loop of area ($\delta
x^{\gamma\psi})$, its length variation is $\delta
y=yV_{\gamma\psi}\delta x^{\gamma\psi}$, where
\begin{equation}\label{1}
V_{\gamma\psi}=\nabla_{\psi}v_{\gamma}-\nabla_{\gamma}v_{\psi},
\end{equation}
and $\nabla _{\psi}$ is the usual covariant derivative. A local
scaling length represented as $\tilde{y}=\sigma(x)y$ induces a
transformation in the vector field $v_{\gamma}$ to
$\tilde{v}_{\gamma}=v_{\gamma}+(\ln\sigma)_{,\gamma}$, where the
metric components are transformed by the conformal transformations
$\tilde{g}_{{\gamma\psi}}=\sigma^{2}g_{\gamma\psi}$ and
$\tilde{g}^{{\gamma\psi}}=\sigma^{-2} g^{\gamma\psi}$.
\begin{equation}\label{2}
{\bar{\Gamma}}^{\lambda}_{\gamma\psi}=\Gamma^{\lambda}_{\gamma\psi}
+g_{\gamma\psi}v^{\lambda}-\delta^{\lambda}_{\gamma}v_{\psi}
-\delta^{\lambda}_{\psi}v_{\gamma},
\end{equation}
where $\Gamma^{\lambda}_{\gamma\psi}$ is the Christoffel symbol. In
Weyl geometry, $\bar{\Gamma}^{\lambda}_{\gamma\psi}$ is symmetric in
its lower indices \cite{40} which helps to evaluate gauge covariant
derivative. The Weyl curvature tensor can be found through the
covariant derivative as
\begin{equation}\label{3}
\mathbb{R}_{\gamma\psi\mu\nu}=\mathbb{R}_{[\gamma\psi]\mu\nu}+
\mathbb{R}_{(\gamma\psi)\mu\nu},
\end{equation}
where
\begin{eqnarray}\label{4}
\mathbb{R}_{(\gamma\psi)\mu\nu}&=&\frac{1}{2}(\mathbb{R}_{\gamma\psi\mu\nu}
+\mathbb{R}_{\psi\gamma\mu\nu})=g_{\gamma\psi}V_{\mu\nu},
\\\nonumber
\mathbb{R}_{[\gamma\psi]\mu\nu}&=&R_{\gamma\psi\mu\nu}
+2\nabla_{\mu}v_{[{\gamma}}g_{\psi]\nu} + 2\nabla_{\nu}v_{[{\psi}}
g_{\gamma]\mu}
\\\label{5}
&+&2v_{\mu}v_{[{\gamma}} g_{\psi]\nu}+2v_{\nu}v_{[{\psi}}
g_{\gamma]\mu}-2v^{2}g_{\mu{[{\gamma}}} g_{\psi]\nu}.
\end{eqnarray}
The contraction of the Weyl curvature tensor gives
\begin{equation}\label{6}
\mathbb{R}^{\gamma}_{\;\psi}=\mathbb{R}^{\mu\gamma}_{\;\mu\psi}
=R^{\gamma}_{\;\psi}+2v^{\gamma}v_{\psi}+3\nabla_{\psi}v^{\gamma}
-\nabla_{\gamma}v^{\psi}+g^{\gamma}_{\;\psi}(\nabla_{\mu}v^{\mu}-2v_{\mu}v^{\mu}),
\end{equation}
where $R^{\gamma}_{\;\psi}$ is the Ricci tensor and the Weyl scalar
is
\begin{equation}\label{7}
\mathbb{R}={R}^{\mu}_{\;\mu}=R+6(\nabla_{\gamma}v^{\gamma}-v_{\gamma}v^{\gamma}).
\end{equation}

A generalization of the Weyl geometry with torsion yields the
Weyl-Cartan (WC) spaces. In this spacetime, one can define a
symmetric metric tensor $g_{\gamma\psi}$ and an asymmetric
connection $\bar{\Gamma}^{\lambda}_{\gamma\psi}$ \cite{41}, where
the torsion tensor can be written as
\begin{equation}\label{8}
\hat{\Gamma}^{\lambda}_{\gamma\psi}={\Gamma}^{\lambda}_{\gamma\psi}
+\mathbb{C}^{\lambda}_{\;\gamma\psi}+\mathbb{L}^{\lambda}_{\;\gamma\psi}.
\end{equation}
Here, $\mathbb{C}^{\lambda}_{\;\gamma\psi}$ is the contortion
tensor, $\mathbb{L}^{\lambda}_{\;\gamma\psi}$ is the disformation
tensor and the Levi-Civita connection is expressed as
\begin{equation}\label{9}
\Gamma^{\lambda}_{\gamma\psi}=\frac{1}{2}g^{\lambda\sigma}
(g_{\sigma\psi,\gamma}+g_{\sigma\gamma,\psi}-g_{\gamma\psi,\sigma}).
\end{equation}
The contortion tensor is defined as
\begin{equation}\label{10}
\mathbb{C}^{\lambda}_{\;\gamma\psi}=\hat{\Gamma}^{\lambda}_{[\gamma\psi]}
+g^{\lambda\sigma}g_{\gamma\kappa}\hat{\Gamma}^{\kappa}_{[\psi\sigma]}
+g^{\lambda\sigma}g_{\psi\kappa}\hat{\Gamma}^{\kappa}_{[\gamma\sigma]},
\end{equation}
which is antisymmetric and
$\hat{\Gamma}^{\lambda}_{[\gamma\psi]}=\frac{1}{2}(\hat{\Gamma}^{\lambda}_{\gamma\psi}
-\hat{\Gamma}^{\lambda}_{\psi\gamma})$. The non-metricity yields the
disformation tensor as
\begin{equation}\label{11}
\mathbb{L}^{\lambda}_{\;\gamma\psi}=\frac{1}{2}g^{\lambda\sigma}(Q_{\psi\gamma\sigma}
+Q_{\gamma\psi\sigma}-Q_{\lambda\gamma\psi}).
\end{equation}
As the negative covariant derivative of the metric tensor with WC
connection $\hat{\Gamma}^{\lambda}_{\gamma\psi}$ is expressed as
$\nabla_{\sigma}g_{\gamma\psi}=Q_{\sigma\gamma\psi}$, the
non-metricity tensor $Q_{\psi\gamma\sigma}$ can be obtained
\begin{equation}\label{12}
Q_{\lambda\gamma\psi}=-g_{\gamma\psi,\lambda}+g_{\psi\sigma}
\hat{\Gamma}^{\sigma}_{\gamma\lambda}
+g_{\sigma\gamma}\hat{\Gamma}^{\sigma}_{\psi\lambda}.
\end{equation}
By comparing Eqs.\eqref{2} and \eqref{8}, we can deduce that Weyl
geometry is a specific instance of WC geometry when the torsion is
zero, and $Q_{\lambda\gamma\psi}=-2g_{\gamma\psi}v_{\lambda}$.
Consequently, the connection in WC geometry can be expressed as
\begin{equation}\label{13}
\hat{\Gamma}^{\lambda}_{\gamma\psi}={\Gamma}^{\lambda}_{\gamma\psi}
+g_{\gamma\psi}v^{\lambda}-\delta^{\lambda}_{\gamma}v_{\psi}
-\delta^{\lambda}_{\psi}v_{\gamma}+\mathbb{C}^{\lambda}_{\;\gamma\psi},
\end{equation}
where
\begin{equation}\label{14}
\mathbb{C}^{\lambda}_{\;\gamma\psi}=T^{\lambda}_{\;\gamma\psi}
-g^{\lambda\nu}g_{\sigma\gamma}T^{\sigma}_{\;\nu\psi}
-g^{\lambda\nu}g_{\sigma\psi}T^{\sigma}_{\;\nu\gamma}.
\end{equation}
The WC torsion is described as
\begin{equation}\label{15}
T^{\lambda}_{\;\gamma\psi}=\frac{1}{2}(\hat{\Gamma}^{\lambda}_{\gamma\psi}
-\hat{\Gamma}^{\lambda}_{\psi\gamma}),
\end{equation}
and the WC curvature tensor as
\begin{equation}\label{16}
\hat{R}^{\lambda}_{\;\gamma\psi\sigma}=\hat{\Gamma}^{\lambda}_{\gamma\sigma,\psi}
-\hat{\Gamma}^{\lambda}_{\gamma\psi,\sigma}+\hat{\Gamma}^{\mu}_{\gamma\sigma}
\hat{\Gamma}^{\lambda}_{\mu\psi}
-\hat{\Gamma}^{\mu}_{\gamma\psi}\hat{\Gamma}^{\lambda}_{\mu\sigma}.
\end{equation}
We can calculate $\hat{R}^{\lambda}_{\;\gamma\psi\sigma}$ in terms
of the Riemann tensor using Eq.\eqref{13} as
\begin{eqnarray}\nonumber
\hat{R}&=&\hat{R}^{\gamma\psi}_{\;~~\gamma\psi}
=R+6\nabla_{\psi}v^{\psi}-4\nabla_{\psi}T^{\psi}-6v_{\psi}v^{\psi}
+8v_{\psi}T^{\psi}+T^{\gamma\mu\psi}T_{\gamma\mu\psi}
\\\label{17}
&+&2T^{\gamma\mu\psi}T_{\psi\mu\gamma}-4T^{\psi}T_{\psi},
\end{eqnarray}
where $T_\gamma=T^\psi_{~\gamma\psi}$, and the covariant derivative
is calculated in relation to the metric tensor.

In the framework of symmetric connection, the Levi-Civita connection
can be represented in relation to the disformation tensor as
\begin{equation}\label{18}
\Gamma^{\lambda}_{\gamma\psi}=-\mathbb{L}^{\lambda}_{\;\gamma\psi}.
\end{equation}
It is well-known from GR that the gravitational action can be
restated in a non-covariant form after removing the boundary terms
from the Ricci scalar as
\begin{equation}\label{19}
S=\frac{1}{2k}\int g^{\gamma\psi}(\Gamma^{\mu}_{\sigma\gamma}
\Gamma^{\sigma}_{\psi\mu} -\Gamma^{\mu}_{\sigma\mu}
\Gamma^{\sigma}_{\gamma\psi})\sqrt{-g} d^ {4}x.
\end{equation}
Using Eq.\eqref{18}, the gravitational action (in the coincident
gauge) in terms of the disformation tensor becomes
\begin{equation}\label{20}
S=-\frac{1}{2k} \int g^{\gamma\psi}(\mathbb{L}^{\mu}_{~\sigma\gamma}
\mathbb{L}^{\sigma}_{~\psi\mu} - \mathbb{L}^{\mu}_{~\sigma\mu}
\mathbb{L}^{\sigma}_{~\gamma\psi}) \sqrt{-g} d^ {4}x.
\end{equation}
This is the action of the STG with some basic differences between
the actions of GR and STG. Firstly, the curvature tensor vanishes in
the STG, thus the overall geometry is characterized by flatness and
it turns out to be Weitzenb$\ddot{o}$ck type geometry \cite{42}.
Secondly, the gravitational effects arising from the variations in
the length of the vector instead of the rotation of the angle
between two vectors during parallel transport.

\subsection{The Field Equations of $f(Q)$ Gravity}

Here, we examine an extension of the STG Lagrangian \eqref{20} as
\begin{equation}\label{21}
S=\int\left(\frac{1}{2k}f(Q)+L_{m}\right) \sqrt{-g}d^{4}x,
\end{equation}
where, determinant of the metric tensor and matter-Lagrangian
density are represented by $g$ and $L_{m}$, respectively. The
non-metricity scalar $Q$ is defined as
\begin{equation}\label{22}
Q=-g^{\gamma\psi}(\mathbb{L}^{\mu}_{~\nu\gamma}\mathbb{L}^{\nu}_{~\psi\mu}
-\mathbb{L}^{\mu}_{~\nu\mu}\mathbb{L}^{\nu}_{~\gamma\psi}),
\end{equation}
while Eq.\eqref{18} gives
\begin{equation}\label{23}
\mathbb{L}^{\mu}_{\;\nu\varsigma}=-\frac{1}{2}g^{\mu\lambda}
(\nabla_{\varsigma}g_{\nu\lambda}+\nabla_{\nu}g_{\lambda\varsigma}
-\nabla_{\lambda}g_{\nu\varsigma}).
\end{equation}
The trace of the $Q$ tensor is given as
\begin{equation}\label{24}
Q_{\mu}=Q^{~\gamma}_{\mu~\gamma},\quad
\tilde{Q}_{\mu}=Q^{\gamma}_{~\mu\gamma}.
\end{equation}
The superpotential is defined as a function of the $Q$ tensor as
\begin{equation}\label{25}
\mathbb{P}^{\mu}_{\;\gamma\psi}=-\frac{1}{2}\mathbb{L}^{\mu}_{\;\gamma\psi}
+\frac{1}{4}(Q^{\mu}-\tilde{Q}^{\mu})g_{\;\gamma\psi}- \frac{1}{4}
\delta ^{\mu}\;_{({\gamma}}Q_{\psi)}.
\end{equation}
Consequently, the relation for $Q$ becomes (details are given in
Appendix \textbf{A})
\begin{equation}\label{26}
Q=-Q_{\mu\gamma\psi}\mathbb{P}^{\mu\gamma\psi}=-\frac{1}{4}(-Q^{\mu\psi\rho}Q_{\mu\psi\rho}
+2Q^{\mu\psi\rho}Q_{\rho\mu\psi}-2Q^{\rho}\tilde{Q}_{\rho}+Q^{\rho}Q_{\rho}).
\end{equation}
The field equations can be obtained by taking the variation of $S$
with respect to the metric tensor as zero
\begin{eqnarray}\nonumber
\delta S=&0=&\int\bigg(\frac{1}{2k}\delta[f(Q)\sqrt{-g}]+\delta
[L_{M}\sqrt{-g}]\bigg)d^{4}x\\\nonumber
&0=&\int\frac{1}{2k}\bigg(\frac{-1}{2}f
g_{\gamma\psi}\sqrt{-g}\delta g^{\gamma\psi}-f_{Q}\sqrt{-g}
(P_{\gamma\mu\nu} Q_{\psi}^{~\mu\nu}- 2Q^{\mu\nu}_{~~~\gamma}
P_{\mu\nu\psi})\delta g^{\gamma\psi}
\\\label{28}
&+&2f_{Q} \sqrt{-g} P_{\mu\gamma\psi} \nabla^{\mu} \delta
g^{\gamma\psi}+\delta [L_{M}\sqrt{-g}]\bigg)d^{4}x,
\end{eqnarray}
where we have used Appendix \textbf{B}. Using the EMT
\begin{equation}\label{29}
T_{\gamma\psi} \equiv \frac{-2}{\sqrt{-g}} \frac{\delta (\sqrt{-g}
L_{M})}{\delta g^{\gamma\psi}},
\end{equation}
and $2 f_{Q}\sqrt{-g}P_{\mu\gamma\psi} \nabla^{\mu}\delta
g^{\gamma\psi}=-2\nabla^{\mu} (f_{Q} \sqrt{-g}
P_{\mu\gamma\psi})\delta g^{\gamma\psi}$, the field equations of
$f(Q)$ gravity take the form
\begin{equation}\label{30}
\frac{-2}{\sqrt{-g}}\nabla_{\gamma}(f_{Q}\sqrt{-g}
P^{\mu}_{~\gamma\psi})-\frac{1}{2}f g_{\gamma\psi}-f_{Q}
(P_{\gamma\mu\nu}Q_{\psi}^{~\mu\nu}-2Q^{\mu\nu}_{~~~\gamma}
P_{\mu\nu\psi})=k^{2} T_{\gamma\psi},
\end{equation}
where $f_{Q}=\frac{\partial f(Q)}{\partial Q}$.

In theoretical cosmology, the most crucial challenge is to
comprehend and regulate the cosmic expansion during both the early
as well as late stages. The idea of modifying matter or gravity has
attracted a great deal of attention. This noteworthy concept of
theoretical advancements, including higher-order gravitational
theories, encompass anti-gravity effects arising from higher-order
curvature terms. A significant advantage of the $f(Q)$ gravity
theory lies in the fact that its field equations are of second
order. Within this framework, the gravitational field is
characterized by the function of non-metricity scalar $Q$.
Non-metricity is a geometric quantity related to the deviation from
metric compatibility in the theory of connections. The specific form
of the function $f(Q)$ determines the gravitational dynamics. This
theory involves non-metricity scalar, introducing additional
geometric structures beyond the metric tensor. We have employed a
recently introduced GGDE model, characterized by a strong repulsive
force that prevents the formation of black holes. These models stand
out as significant solutions to the recent cosmic acceleration
problem.

\section{Reconstruction of GGDE $f(Q)$ Model}

To reconstruct the GGDE $f(Q)$ model, we use a correspondence
technique between GGDE and $f(Q)$ gravity in this section. The line
element of spatially homogeneous and isotropic universe model is
given by
\begin{equation}\label{32}
dS^{2}=-dt^{2}+a^{2}(t)(dx^{2}+dy^{2}+dz^{2}),
\end{equation}
where the scale factor is denoted by $a$. The EMT for perfect fluid
configuration containing four velocity field $u_{\gamma}$, the usual
matter density and pressure $\rho_{m}$ and $p_{m}$, respectively, is
given as
\begin{equation}\label{33}
\tilde{T}_{\;\gamma\psi}=(\rho_{m}+p_{m})u_{\gamma}u_{\psi}+p_{m}g_{\gamma\psi},
\end{equation}
The modified Friedman equations for $f(Q)$ gravity are
\begin{equation}\label{34}
3H^{2}=\rho_{m}+\rho_D,\quad 2\dot{H}+3H^{2}=p_{m}+p_D,
\end{equation}
here dot means derivative with respect to time while $\rho_D$ and
$p_D$ are the DE density and pressure, respectively, given by
\begin{eqnarray}\label{35}
\rho_D&=&\frac{f}{2}-6H^{2}f_{Q},\\\label{36}
p_D&=&\frac{f}{2}+2f_{Q}\dot{H}+2Hf_{QQ}+6H^{2}f_{Q},
\end{eqnarray}
The two fractional energy densities $\Omega_{D}$ and $\Omega_{m}$
are given as follows
\begin{equation}\label{37}
\Omega_{D}=\frac{\rho_{D}}{\rho_{cr}}=\frac{\rho_{D}}{3H^{2}}, \quad
\Omega_{m}=\frac{\rho_{m}}{\rho_{cr}}=\frac{\rho_{m}}{3H^{2}},
\end{equation}
implying that $1=\Omega_{D}+\Omega_{m}$, where $\rho_{cr}$ is the
critical density. Consider the interaction between two fluids, i.e.,
the DE and dark matter. Therefore, the continuity of two fluids
yields that the energy densities of the two fluids do not conserve
separately but takes the following form for the interacting case
\begin{equation}\label{38}
\dot{\rho}_{m}+3H(\rho_{m}+p_{m})=\Gamma,\quad
\dot{\rho}_{D}+3H(\rho_{D}+p_{D})=-\Gamma,
\end{equation}
here, $\Gamma$ represents the interaction term. It is evident that
$\Gamma$ must be positive, signifying the occurrence of energy
transfer from DE to DM. Since the unit of $\Gamma$ is the inverse of
time evolution, it is natural to choose this value as the product of
$H$ and $\rho_{D}$. In this context, we examine $\Gamma = 3\eta
H(\rho_{m}+p_{D}) = 3\eta H\rho_{D}(1+u)$, where $\eta$ denotes the
coupling constant, representing the transfer strength from DE to DM
and is given as
\begin{equation}\label{40}
u=\frac{\rho_{m}}{\rho_{D}}=\frac{\Omega_{m}}{\Omega_{D}}
=\frac{1-\Omega_{D}}{\Omega_{D}}.
\end{equation}
With the parameters defined above, we can write $\omega_{D}$ as
\begin{equation}\label{41}
\omega_{D}=-\frac{1}{2-\Omega_{D}}\bigg(1+\frac{2\eta}{\Omega_{D}}\bigg).
\end{equation}

Dynamical DE models involving energy density proportional to the
Hubble parameter are essential to explain the accelerated expansion
of the universe. In this scenario, the GGDE model is a dynamical DE
model possessing energy density as The energy density for GGDE model
can be expressed as
\begin{equation}\label{42}
\rho_D=\xi H+\zeta H^{2}.
\end{equation}
The $f(Q)$ theory has successfully undergone observational tests
within the confines of the solar system \cite{14a}. Considering the
dynamic behavior of the GGDE model, it is more appropriate to
analyze this model within a dynamic framework, such as the $f(Q)$
theory. It has been demonstrated that certain characteristics of the
original GDE in $f(Q)$ cosmology deviate from GR. A phenomenological
model of GGDE was recently introduced, characterized by an energy
density expressed as $\rho_D=\xi H+\zeta H^{2}$. This model aims to
provide an explanation for the observed acceleration of the universe
expansion. The subleading term $H^{2}$ can induce delays in various
epochs of cosmic evolution. The GGDE model mimics the behavior of
cosmological constant. It is noteworthy that this behavior is
consistent with that of the original GDE model. This is expected, as
the $H^2$ in the late time can be neglected due to the smallness of
$H$, and the distinction between these two models becomes apparent
only in the early epochs of the universe. The solution of the
Friedman equations of our model leads to a stable universe in $f(Q)$
theory due to non-metricity factor $Q$.

By equating the associated densities, we establish the relationship
between GGDE and the $f(Q)$ gravity model \cite{14b}. Using
Eqs.\eqref{35} and \eqref{42}, it is obvious that
\begin{equation}\label{43}
\frac{f}{2}-6H^{2}f_{Q}=\xi H+\zeta H^{2},
\end{equation}
which can also be written as
\begin{equation}\label{44}
f_{Q}-\frac{f}{12H^{2}}+\frac{\xi}{6H}+\frac{\zeta}{6}=0.
\end{equation}
The represents the first-order linear differential equation in terms
of $Q$ whose solution is
\begin{equation}\label{45}
f(Q)=\sqrt{Q}c_{1}+c_{2}Q+c_{3}\sqrt{Q}\ln Q,
\end{equation}
where $c_{1},~c_{2}$ and $c_{3}$ are integration constants. This is
the reconstructed GGDE model in $f(Q)$ gravity.

Now, we write down this solution in terms of the redshift parameter.
We express the scale factor in the form of a power law, given by
\begin{equation}\label{46}
a(t)=a_{0}t^{b},
\end{equation}
where $a_0$ is an arbitrary constant whose current value is $1$ and
$b$ is an arbitrary constant. The deceleration parameter is defined
as
\begin{equation}\label{47}
q=-\frac{a\ddot{a}}{\dot{a}^{2}}=-1+\frac{1}{b}.
\end{equation}
When $q<0$, it represents that the universe is accelerating while
$q>0$ indicates deceleration. Substituting the value of $b$ in
Eq.\eqref{46}, we obtain
\begin{equation}\label{48}
a(t)=t^{\frac{1}{1+q}},
\end{equation}
where $q>-1$ represents the expanding universe. Moreover, the
current universe is accelerating, as indicated by the current value
of the deceleration parameter, i.e., $q=-0.832^{+0.091}_{-0.091}$
\cite{44b}. The historical expansion and the current expansion rates
of the universe can be expressed as
\begin{equation}\label{49}
H=\frac{\dot{a}}{a}=\left(\frac{1}{1+q}\right)t^{-1},\quad
H_0=\left(\frac{1}{1+q}\right)t_{0}^{-1}.
\end{equation}
This indicates that the parameters $q$ and $H_0$ determine the
cosmic expansion. Using the relationship between the scale factor
and the redshift parameter $z~(a=\Upsilon^{-1})$, we have
\begin{equation}\label{50}
H=H_0\Upsilon^{1+q} , \quad \dot{H} =-H_0\Upsilon^{2+2q},
\end{equation}
where $\Upsilon=1+z$.
\begin{figure}
\epsfig{file=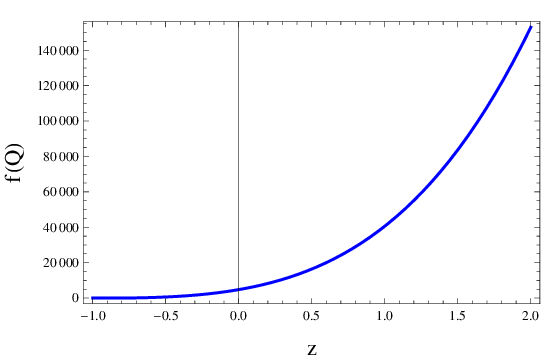,width=0.5\linewidth}
\epsfig{file=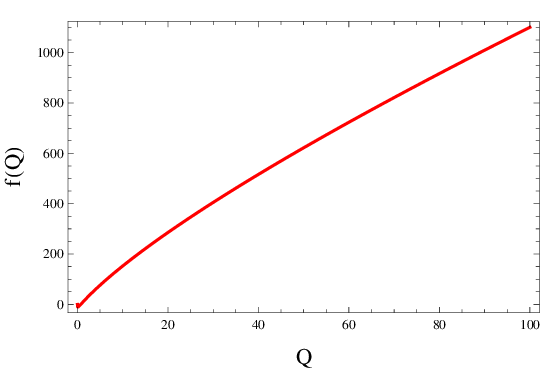,width=0.5\linewidth}\caption{Graphical
representation of $f(Q)$ versus $z$ and $Q$.}
\end{figure}

One can calculate $Q$ as $Q=6H^{2}$ (details are given in Appendix
\textbf{B}). Using the value of $H$, we have
\begin{equation}\label{51}
Q=6H_0^{2}\Upsilon^{2+2q}.
\end{equation}
Inserting this value in Eq.\eqref{45}, the solution in terms of
redshift parameter can be written as
\begin{equation}\label{52}
f(Q)=\sqrt{6H_0^{2}\Upsilon^{2+2q}}c_{1}+6c_{2}H_0^{2}\Upsilon^{2+2q}
+c_{3}\sqrt{6H_0^{2}\Upsilon^{2+2q}}\ln (6H_0^{2}\Upsilon^{2+2q}).
\end{equation}
We take $c_1=1,~c_2=-4,~c_3=-15$ throughout the graphical analysis.
Figure \textbf{1} shows that the reconstructed GGDE model maintains
positive value and increases with respect to both $z$ as well as
$Q$. This analysis suggests that the GGDE model signifies an
accelerated expansion. We also note that our reconstructed model
approaches to zero when $Q$ tends to zero which implies a realistic
behavior of the resulting model.
\begin{figure}
\epsfig{file=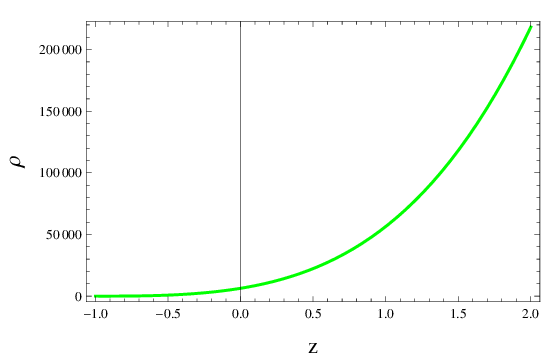,width=.5\linewidth}
\epsfig{file=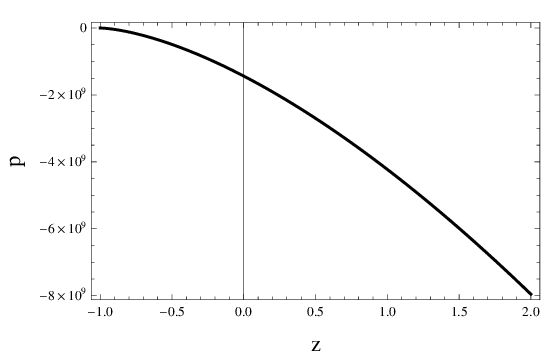,width=.5\linewidth}\caption{Plots of $\rho_D$
and $p_D$ versus $z$.}
\end{figure}

Next, we investigate the behavior of $\rho_D$ and $p_D$ for the
reconstructed GGDE $f(Q)$ gravity model. Using Eq.\eqref{45} in
\eqref{35} and \eqref{36}, it follows that
\begin{eqnarray}\label{53}
\rho_D&=&\frac{1}{2}\sqrt{Q}(2c_{1}+2c_{3}+3c_{2}\sqrt{Q}+2c_{3}\ln
Q),\\\nonumber
p_D&=&2(\dot{H}+3H^{2})\bigg(c_{2}+\frac{c_{1}}{2\sqrt{Q}}
+\frac{c_{3}}{\sqrt{Q}}+\frac{c_{3}\ln
Q}{2\sqrt{Q}}\bigg)\\\label{54}&-&2H\dot{Q}
\bigg(\frac{c_{1}}{4Q^{\frac{3}{2}}}+\frac{c_{3}\ln
Q}{{4Q^{\frac{3}{2}}}}\bigg)-\frac{\sqrt{Q}c_{1}+c_{2}Q+c_{3}
\sqrt{Q}\ln Q}{2}.
\end{eqnarray}
Converting these equations in terms of redshift parameter, we have
\begin{equation}\label{55}
\rho_D=\bigg(2c_{1}+2c_{3}+3c_{2}\sqrt{6H_0^{2}\Upsilon^{2+2q}}+2c_{3}\ln(6
H_0^{2}\Upsilon^{2+2q})\bigg)\sqrt{\frac{3}{2}H_0^{2}\Upsilon^{2+2q}},
\end{equation}
\begin{eqnarray}\nonumber
p_D&=&\frac{-H_0\Upsilon^{1+q}}{12\sqrt{6}(H_0^{2}\Upsilon^{2+2q})^{\frac{3}{2}}}
\bigg(c_{1}(1+12H_0^{2}\Upsilon^{3+3q})+12H_0^{2}\Upsilon^{3+3q}\\
\nonumber
&+&\bigg(c_{3}(-2+6H_0)+c_{2}(-2+3H_0)\sqrt{6H_0^{2}\Upsilon^{2+2q}}\bigg)
\\\label{56}
&-&c_{3}(1+12H_0^{2}\Upsilon^{3+3q})\ln(6
H_0^{2}\Upsilon^{2+2q})\bigg).
\end{eqnarray}
Figures \textbf{2} shows the behavior of $\rho_D$ and $p_D$ with
redshift parameter. The energy density $\rho_D$ maintains positive
behavior and increases, while the $p_D$ maintains negative behavior
and exhibits a decreasing trend for all $z>-1$ which is consistent
with the DE behavior.

\section{Cosmographic Analysis}

In this section, we investigate cosmic evolution through
cosmographic analysis of the EoS parameter and phase planes for the
reconstructed GGDE $f(Q)$ model in the interacting case. We also
discuss the stability of this model by considering $\nu_{s}^{2}$.

\subsection{Equation of State Parameter}

In the field of cosmology, the EoS parameter
($\omega_{D}=\frac{p_{D}}{\rho_{D}}$) can be classified into various
phases of the cosmic evolution. The values of
$\omega_{D}=0,~\frac{1}{3},~1$ indicate the matter-dominated
regions, i.e., dust, radiation and stiff matter, respectively. For
the DE era, $\omega_{D}=
-1,~\omega_{D}<-1,~-1<\omega_{D}<-\frac{1}{3}$ lead to vacuum,
phantom and quintessence phases of the universe expansion. Using
Eq.\eqref{41}, we have
\begin{eqnarray}\nonumber
\omega_{D}&=&\bigg[\sqrt{Q}\bigg(2\eta+\frac{2c_{1}-2c_{3}-3c_{2}\sqrt{Q}
- 2c_{3}\ln Q}{\sqrt{Q}}\bigg)\bigg]\\\nonumber
&\times&\bigg[\bigg(-2c_{1}+2c_{3}+3c_{2}\sqrt{Q}+2c_{3}\ln Q
\bigg)\\\label{57}
&\times&\bigg(2+\frac{-2c_{1}+2c_{3}+3c_{2}\sqrt{Q}+2c_{3}\ln
Q}{\sqrt{Q}}\bigg)\bigg]^{-1}.
\end{eqnarray}
This equation in terms of the redshift parameter becomes
\begin{eqnarray}\nonumber
\omega_{D}&=&\bigg[\bigg\{\sqrt{6}\sqrt{H_0^{2}\Upsilon^{2+2q}}\bigg\}
\bigg\{2\eta-\bigg(2c_{3}\ln(6H_0^{2}\Upsilon^{2+2q})\\
\nonumber
&-&2c_{1}+2c_{3}+3c_{2}\sqrt{6H_0^{2}\Upsilon^{2+2q}}\bigg)\bigg(\sqrt{6H_0^{2}
\Upsilon^{2+2q}}\bigg)^{-1}\bigg\}\bigg]\\
\nonumber
&\times&\bigg[\bigg\{2c_{3}-2c_{1}+3c_{2}\sqrt{6H_0^{2}\Upsilon^{2+2q}}+2c_{3}
\ln(6H_0^{2}\Upsilon^{2+2q})\bigg\}\\
\nonumber
&\times&\bigg\{\bigg(\big(2c_{3}-2c_{1}+3c_{2}\sqrt{6H_0^{2}
\Upsilon^{2+2q}}+2c_{3}\ln(6H_0^{2}\Upsilon^{2+2q})\big)\\
\label{58} &\times&
\bigg(\sqrt{6H_0^{2}\Upsilon^{2+2q}}~\bigg)^{-1}\bigg)+2\bigg\}\bigg]^{-1}.
\end{eqnarray}
Figure \textbf{3} illustrates the dynamic evolution of the EoS in
the GGDE $f(Q)$ gravity model for three different values of
$\eta=10.3,~ 10.4,~ 10.5$. This indicates that $\omega_{D}<-1$ in
the late universe ($z>0$), yielding a phantom field DE. This phantom
phase is maintained for $\eta>10.5$, whereas this represents a
quintessence phase ($\omega_{D}>-1$) for $\eta<10.3$.
\begin{figure}[H]\center
\epsfig{file=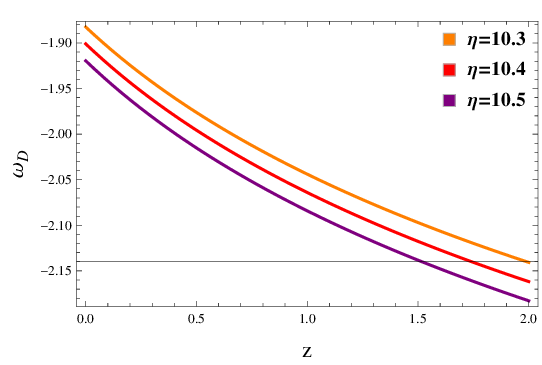,width=.6\linewidth}\caption{Graph of
$\omega_{D}$ with $z$.}
\end{figure}

\subsection{The ($\omega_{D}-\omega^{\prime}_{D}$)-Plane}

Caldwell and Eric \cite{46} examined the behavior of quintessence
scalar field DE models. They analyzed the
$(\omega_{D}-\omega^{\prime}_{D})$-plane, where
$\omega^{\prime}_{D}$ represents the rate of change or evolution of
$\omega_{D}$ with respect to $Q$. We can categorize the models into
two distinct classes, i.e., thawing and freezing regions. The
thawing region is characterized by
$\omega_{D}<0,~\omega^{\prime}_{D}>0$, while the freezing region by
$\omega_{D}<0,~\omega^{\prime}_{D}<0$. Differentiating Eq.
\eqref{57} with respect to $Q$, we get
\begin{eqnarray}\nonumber
\omega^{\prime}_{D}&=&-\bigg[\bigg\{8c_{3}\sqrt{Q}\bigg\}
\bigg\{\big(2c_{1}-2c_{3}-8\sqrt{Q}-3c_{2}\sqrt{Q}-2c_{3}\ln Q\big)\\
\nonumber &\times&\big(2c_{1}-2c_{3}-3c_{2}\sqrt{Q}-2c_{3}\ln
Q\big)\bigg\}^{-1}\bigg]
-\bigg[\bigg\{\big(2c_{3}+3c_{2}\sqrt{Q}\\
\nonumber &-&2c_{1}-8\sqrt{Q}\eta+2c_{3}\ln
Q\big)8c_{3}\sqrt{Q}\bigg\}
\bigg\{\big(2c_{1}-2c_{3}-3c_{2}\sqrt{Q}\\
\nonumber &-&2c_{3}\ln
Q\big)\big(2c_{1}-2c_{3}-8\sqrt{Q}-3c_{2}\sqrt{Q}-2c_{3}\ln Q\big)^{2}\bigg\}^{-1}\bigg]\\
\nonumber&-&\bigg[\bigg\{8c_{3}\sqrt{Q}\big(-2c_{1}+2c_{3}+3c_{2}
\sqrt{Q}-8\sqrt{Q}\eta+2c_{3}\ln Q\big)\bigg\}\\
\nonumber
&\times&\bigg\{\big(2c_{1}-2c_{3}-8\sqrt{Q}-3c_{2}\sqrt{Q}-2c_{3}\ln
Q\big)\big(2c_{1}-3c_{2}\sqrt{Q}
\\\nonumber
&-&2c_{3}\ln Q-2c_{3}\big)^{2}\bigg\}^{-1}\bigg].
\end{eqnarray}
In terms of redshift parameter, this takes the form
\begin{eqnarray}\nonumber
\omega^{\prime}_{D}&=&\bigg[16\sqrt{6} c_{3}\sqrt{H_0^{2}\Upsilon^{2+2q}}
\bigg\{2c_{1}^{2}+2c_{3}^{2}+27c_{2}^{2} H_0^{2}\Upsilon^{2q}\\
\nonumber &+& 54c_{2}^{2} H_0^{2}z\Upsilon^{2q}+27c_{2}^{2} H_0^{2}
z^{2}\Upsilon^{2q}-192H_0^{2}\Upsilon^{2q}\eta \\
\nonumber &-&
144c_{2}H_0^{2}\Upsilon^{2q}\eta-2c_{1}\big(2c_{3}+\sqrt{6H_0^{2}\Upsilon^{2+2q}}
(3c_{2}-8\eta)\big)\\
\nonumber &+&6\sqrt{6}c_{2}c_{3}\sqrt{H_0^{2}\Upsilon^{2+2q}}
-288c_{2}H_0^{2}z\Upsilon^{2q}\eta\\
\nonumber
&-&384H_0^{2}z\Upsilon^{2q}\eta-192H_0^{2}z^{2}\Upsilon^{2q}\eta-
144c_{2}H_0^{2}z^{2}\Upsilon^{2q}\eta\\
\nonumber &-& 16 \sqrt{6} c_{3}\sqrt{H_0^{2}\Upsilon^{2+2q}}\eta+2c_{3}
\ln(6H_0^{2}\Upsilon^{2+2q})\big(2c_{3}-2c_{1}\\
\nonumber
&+&\sqrt{6}\sqrt{H_0^{2}\Upsilon^{2+2q}}\big(3c_{2}-8\eta\big)\big)+2c_{3}^{2}
\ln\big(6H_0^{2}\Upsilon^{2+2q}\big)^{2}\bigg\}\bigg]  \\
\nonumber &\times&\bigg[\bigg\{\big(2c_{3}-2c_{1}
+3c_{2}\sqrt{6H_0^{2}\Upsilon^{2+2q}}+2c_{3}\ln(6H_0^{2}\Upsilon^{2+2q})\big)^{2}\\
\nonumber&\times&\big(-2c_{1}+2c_{3}+8\sqrt{6}\sqrt{H_0^{2}\Upsilon^{2+2
q}}+3\sqrt{6} c_{2}\sqrt{H_0^{2}\Upsilon^{2+2 q}}\\
\label{60}&+&2c_{3}\ln
(6H_0^{2}\Upsilon^{2+2q})\big)^{2}\bigg\}\bigg]^{-1}.
\end{eqnarray}
Figure \textbf{4} shows that $\omega_{D}<0,~\omega^{\prime}_{D}<0$
for all values of $\eta$, which ensures the presence of the freezing
region. This indicates that the cosmological expansion is
accelerating at a higher rate within this framework.

Here, we utilize the $(\omega_{D}-\omega^{\prime}_{D})$-plane in
order to illustrate the current cosmic expansion paradigm. In our
work, behavior of the $(\omega_{D}-\omega^{\prime}_{D})$-plane
represents freezing region in interacting GGDE model within the
framework of $f(Q)$ gravity. However, the result depicts that
$(\omega_{D}-\omega^{\prime}_{D})$-plane represents thawing region
which is less accelerating phase as compared with freezing region in
$f(G)$ gravity \cite{38b}.
\begin{figure}[H]\center
\epsfig{file=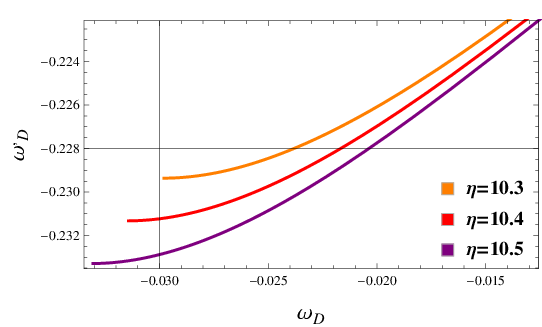,width=.6\linewidth}\caption{Plot of
$\omega^{\prime}_{D}$ versus $\omega_{D}$.}
\end{figure}

\subsection{The $(r-s)$-Plane}

Sahni et al. \cite{47} presented a set of dimensionless parameters
$(r, s)$, which are referred to as the statefinder parameters. These
are characterized as \cite{1d}
\begin{equation}\nonumber
r=\frac{\dddot{a}}{aH^{3}}, \quad s=\frac{r-1}{3(q-\frac{1}{2})},
\end{equation}
In this scenario, the universe comprises two distinct components of
the EoS parameter: the usual matter represented by $\omega_{m}$ and
an exotic form of energy denoted as $\omega_{D}$. The $(r,s)$
parameters can be defined as \cite{1d}
\begin{equation}\nonumber
r=1+\frac{9\omega_{D}}{2}\Omega_{D}(1+\omega_{D})
-\frac{3{\omega'}_{D}}{2H}\Omega_{D},\quad
s=1+\omega_{D}-\frac{{\omega'}_{D}}{3\omega_{D}H}.
\end{equation}
These are helpful in distinguishing various DE models of the cosmos
and also give distance of a particular DE model through lambda cold
dark matter limit. The literature indicates CDM limit for
$(r;s)=(1;0)$ and lambda CDM limit for $(r;s)=(1;1)$. When the
trajectories of the $(r-s)$ plane lie in the range of $(r<1;s>0)$,
we have phantom and quintessence DE eras and for the range
$(r>1;s<0)$, we obtain Chaplygin gas model. The $(r,s)$ diagnostic
in terms of redshift parameter are given as follows
\begin{eqnarray}\nonumber
r&=&\frac{1}{8} \bigg[8+\bigg[\bigg\{36 \bigg(2c_{1}-2c_{3}-3c_{2}
\sqrt{6H_0^{2}\Upsilon^{2q+2}}-2c_{3}\ln(6H_0^{2}\Upsilon^{2q+2})\\
\nonumber &+&8 \sqrt{6}
\sqrt{H_0^{2}\Upsilon^{2q+2}}\eta\bigg)\bigg(4c_{1}^{2}-4c_{1}\big((3
c_{2}+2)\sqrt{6}\sqrt{H_0^{2}\Upsilon^{2q+2}}+2c_{3}-2c_{1}\big)\\
\nonumber &+&4c_{3}\ln(6H_0^{2}\Upsilon^{2q+2})\big((3c_{2}+2)
\sqrt{6H_0^{2}\Upsilon^{2q+2}}+2c_{3}\big)+54c_{2}^{2}H_0^{2}\\
\nonumber &\times&\Upsilon^{2q+2}+12c_{2}c_{3}\sqrt{6}
\sqrt{H_0^{2}\Upsilon^{2q+2}}+192 \eta
H_0^{2}\Upsilon^{2q+2}+72c_{2}H_0^{2}
\\
\nonumber &\times&\Upsilon^{2q+2}+4c_{3}^{2}
\ln(6H_0^{2}\Upsilon^{2q+2})^{2}+8
c_{3}\sqrt{6H_0^{2}\Upsilon^{2q+2}}+4c_{3}^2\bigg)\bigg\}\\
\nonumber &\times&\bigg\{\bigg(2c_{1}-3
c_{2}\sqrt{6H_0^{2}\Upsilon^{2q+2}}
-2c_{3}\ln(6H_0^{2}\Upsilon^{2q+2})-2c_{3}\bigg)\bigg(8\sqrt{6}H_0\\
\nonumber &\times& \sqrt{\Upsilon^{2q+2}}+2c_{3}\ln(6
H_0^{2}\Upsilon^{2q+2})+2c_{3}+\sqrt{6H_0^{2}\Upsilon^{2q+2}}\bigg)^{2}\bigg\}^{-1}\bigg]\\
\nonumber &+&\bigg[\bigg\{24c_{3}\bigg(-4c_{1}
\bigg(\sqrt{6H_0^{2}\Upsilon^{2q+2}}(3c_{2}-8 \eta )
+2c_{3}\bigg)+4 c_{3}\ln(6H_0^{2}\\
\nonumber
&\times&\Upsilon^{2q+2})+4c_{1}^{3}\bigg(2c_{3}-2c_{1}+\sqrt{6H_0^{2}\Upsilon^{2q+2}}
(3c_{2}-8\eta)\bigg)+12 c_{2}c_{3}\sqrt{6}\\
\nonumber &\times& \sqrt{H_0^{2}\Upsilon^{2q+2}}-288c_{2}\eta
H_0^{2}\Upsilon^{2q+2}+4 c_{3}^{2}
\ln^2(6 H_0^{2}\Upsilon^{2q+2})+4c_{3}^{2}\\
\nonumber &+&54 c_{2}^{2}H_0^{2}\Upsilon^{2q+2}-32 c_{3}
\eta\sqrt{6}\sqrt{H_0^{2}\Upsilon^{2q+2}}-384\eta
H_0^{2}\Upsilon^{2q+2}\bigg)\bigg\}\\
\nonumber &\times&\bigg\{\sqrt{H_0^{2}\Upsilon^{2q+2}}\bigg(3c_{2}
\sqrt{6H_0^{2}\Upsilon^{2q+2}}-2c_{1}+2c_{3}\log(6H_0^{2}\Upsilon^{2q+2})
\\ \nonumber &+&2c_{3}\bigg)\bigg(-2c_{1}+3c_{2}\sqrt{6} \sqrt{H_0^{2}\Upsilon^{2q+2}}
+2c_{3}\ln(6H_0^{2}\Upsilon^{2q+2})+2 c_{3} \\
\label{66} &+&8 \sqrt{6}
\sqrt{H_0^{2}\Upsilon^{2q+2}}\bigg)^2\bigg\}^{-1}\bigg]\bigg],
\\\nonumber
s&=&1-\bigg[\bigg\{4\sqrt{6H_0^{2}\Upsilon^{2q+2}}\bigg(-2c_{1}+3c_{2}\sqrt{6}
\sqrt{H_0^{2}\Upsilon^{2q+2}}+2c_{3}\\
\nonumber &+&2c_{3}\ln(6H_0^{2}\Upsilon^{2q+2})-8\eta\sqrt{6}
\sqrt{H_0^{2}\Upsilon^{2q+2}}\bigg)\bigg\}\bigg\{\bigg(2c_{1}-2c_{3}\\
\nonumber &-&3c_{2}\sqrt{6H_0^{2}\Upsilon^{2q+2}}-2c_{3}\ln(6
H_0^{2}\Upsilon^{2q+2})\bigg)\bigg(2c_{3}-2c_{1}+3c_{2}H_0 \\
\nonumber &\times&\sqrt{6\Upsilon^{2q+2}}+2c_{3}\ln(6
H_0^{2}\Upsilon^{2q+2})+8\sqrt{6H_0^{2}\Upsilon^{2q+2}}\bigg)\bigg\}^{-1}\bigg]\\
\nonumber&+& \bigg[\bigg\{c_{3}\bigg(4c_{1}^{2}-4c_{1}
\bigg(\sqrt{6}
\sqrt{H_0^{2}\Upsilon^{2q+2}}(3c_{2}-8 \eta)+2c_{3}\bigg)+12c_{2} c_{3}\sqrt{6}\\
\nonumber&+&54c_{2}^{2}H_0^{2}\Upsilon^{2q+2}+4
c_{3}^{2}\sqrt{H_0^{2}\Upsilon^{2q+2}}-4c_{3}\ln(6h^{2}\Upsilon^{2q+2})\\
\nonumber&\times&
\bigg(\sqrt{6}\sqrt{H_0^{2}\Upsilon^{2q+2}}(3c_{2}-8\eta )-2c_{1}+2
c_{3}\bigg)-288c_{2}\eta H_0^{2}\Upsilon^{2q+2}\\
\nonumber &+& 32c_{3}\eta \sqrt{6H_0^{2}\Upsilon^{2q+2}}-384 \eta
H_0^{2}\Upsilon^{2q+2}+4 c_{3}^{2}
\ln(6H_0^{2}\Upsilon^{2q+2})^{2}\bigg)\bigg\}\\
\nonumber &\times&
\bigg\{\sqrt{H_0^{2}\Upsilon^{2q+2}}\bigg(3c_{2}\sqrt{6}
\sqrt{H_0^{2}\Upsilon^{2q+2}}+2c_{3}\ln(6H_0^{2}\Upsilon^{2q+2})\\
\nonumber&+& 2c_{3}-2c_{1}\bigg)\bigg(-2c_{1}+3c_{2}\sqrt{6}
\sqrt{H_0^{2}\Upsilon^{2q+2}}+2c_{3}\ln(6H_0^{2}\Upsilon^{2q+2})\\
\nonumber&+& 2c_{3}+8\sqrt{6}
\sqrt{H_0^{2}\Upsilon^{2q+2}}\bigg)\bigg(-2c_{1}+3c_{2}\sqrt{6}
\sqrt{H_0^{2}\Upsilon^{2q+2}}
\\\label{67}
&+& 2c_{3}\ln (6H_0^{2}\Upsilon^{2q+2})+2c_{3}-8 \eta \sqrt{6}
\sqrt{H_0^{2}\Upsilon^{2q+2}}\bigg)\bigg\}^{-1}\bigg].
\end{eqnarray}
Figure \textbf{5} shows that $(r>1;s<0)$, indicating Chaplygin gas
model for $\eta=10.3, 10.4, 10.5$. When $\eta<10.3$, the
trajectories of the $(r-s)$ plane lie in the range of $(r<1;s>0)$,
indicating phantom and quintessence DE eras and for $\eta<10.5$, the
trajectories lie in the range of $(r>1;s<0)$, indicating Chaplygin
gas model.
\begin{figure}\center
\epsfig{file=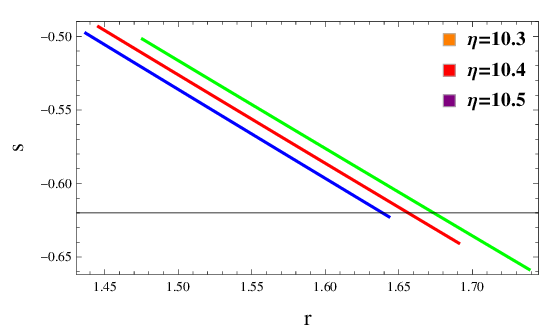,width=.6\linewidth}\caption{Plots of $r$ versus
$s$.}
\end{figure}

The acceleration of cosmic expansion is specified by parameter s
whereas the deviation from pure power-law behavior is specifically
defined by the parameter $r$. It is a geometric diagnostic that does
not support any specific cosmological paradigm. Furthermore, it
attains CDM limit but can not achieve $\Lambda CDM$ limit. We have
observed that, for a significant portion, the $r-s$ plane results in
the phantom region in the context of the Chaplygin gas model.
Furthermore, our reconstructed $f(G)$ model, within the framework of
an interacting scenario, attains the CDM limit. This is in contrast
to the correspondence between $f(G)$ and GGDE model. Our results are
consistent with these cosmos \cite{38b}.

\subsection{The Squared Speed of Sound Parameter}

To assess the stability of the DE model, perturbation theory offers
a direct analysis by examining the sign of the $\nu_{s}^{2}$. When
$\nu_{s}^{2}$ is negative, perturbations intensify, leading to an
unstable state. Conversely, if $\nu_{s}^{2}$ is positive,
perturbations exhibit oscillatory behavior, suggesting the stability
of the background in the presence of linear disturbances. The
corresponding $\nu_{s}^{2}$ is given as
\begin{eqnarray}\nonumber
\nu_{s}^{2}&=&\frac{\dot{p}_{D}}{\dot{\rho}_{D}}
=\frac{\rho_D}{{\dot{\rho}_{D}}}\omega'_{D}+\omega_{D},
\\\nonumber
\nu_{s}^{2}&=&\bigg[\bigg\{-\frac{1}{2}\sqrt{Q}\big(2c_{3}-2c_{1}
+3c_{2}\sqrt{Q}+2c_{3}\ln
Q\big)\bigg\}\bigg\{\frac{1}{2}\bigg(\frac{2 c_{3}}{Q}
-\frac{3c_{2}}{2\sqrt{Q}}\bigg)\\\nonumber
&\times&\sqrt{Q}-\frac{-2c_{1}+2c_{3}+3 c_{2}\sqrt{Q}+2c_{3}\ln
Q}{4\sqrt{Q}}\bigg\}^{-1}\bigg]\bigg[-\bigg[\bigg\{8c_{3}\sqrt{Q}\bigg\}\\
\nonumber &\times&\bigg\{\big(2c_{1}-2c_{3}-3c_{2}\sqrt{Q}-2c_{3}\ln
Q\big)\big(2c_{1}-2c_{3}-8\sqrt{Q}-3c_{2}\sqrt{Q}\\
\nonumber &-&2c_{3}\ln
Q\big)\bigg\}^{-1}\bigg]\bigg[\bigg\{(-2c_{1}+2c_{3}+3c_{2}\sqrt{Q}
-8\sqrt{Q}\eta+2c_{3}\ln Q)\\
\nonumber&\times&8c_{3}\sqrt{Q}\bigg\}\bigg\{(2c_{1}-2c_{3}-3c_{2}\sqrt{Q}-2c_{3}\ln
Q)(2c_{1}-2c_{3}-8\sqrt{Q}\\
\nonumber&-&3c_{2}\sqrt{Q}-2c_{3}\ln
Q)^{2}\bigg\}\bigg]-\bigg[\bigg\{8c_{3}\sqrt{Q}(2c_{3}+3c_{2}\sqrt{Q}
-8\sqrt{Q}\eta\\
\nonumber&-&2c_{1}+2c_{3}\ln
Q)\bigg\}-\bigg\{(2c_{1}-2c_{3}-8\sqrt{Q}-3c_{2}\sqrt{Q}-2c_{3}\ln
Q)\\
\nonumber &\times&(2c_{1}-2c_{3}-3c_{2}\sqrt{Q}-2c_{3}\ln
Q)^{2}\bigg\}^{-1}\bigg]\bigg]+\bigg[\bigg\{\sqrt{Q}\bigg(2\eta\\
\nonumber&+&\frac{2c_{1}-2c_{3}-3c_{2}\sqrt{Q}-2c_{3}\ln
Q}{\sqrt{Q}}\bigg)\bigg\}\bigg\{(2c_{3}-2c_{1}+3c_{2}\sqrt{Q}+2c_{3}\ln
Q)\\\nonumber
&\times&\bigg(2+\frac{2c_{3}-2c_{1}+3c_{2}\sqrt{Q}+2c_{3}\ln
Q}{\sqrt{Q}}\bigg)\bigg\}\bigg].
\end{eqnarray}
In terms of redshift parameter, it becomes
\begin{eqnarray}\nonumber
\nu_{s}^{2}&=&\bigg[\sqrt{6}\sqrt{H_0^{2}\Upsilon^{2+2
q}}\bigg[\bigg\{96c_{3}H_0^{2}\Upsilon^{2+2q}\bigg(27c_{2}^{2}
H_0^{2}\Upsilon^{2q}
\\\nonumber
&+& 2c_{1}^{2}+2c_{3}^{2}+54c_{2}^{2}H_0^{2}\Upsilon^{2q}z +6
\sqrt{6} c_{2} c_{3}\sqrt{H_0^{2}\Upsilon^{2+2q}}\\\nonumber
&+&27c_{2}^{2}H_0^{2}\Upsilon^{2q} z^{2}-2c_{1} \big(2c_{3}
+\sqrt{6H_0^{2}\Upsilon^{2+2q}}(3c_{2}-8\eta)\big)\\\nonumber &-&192
H_0^{2}\Upsilon^{2q}\eta-144 c_{2}H_0^{2}\Upsilon^{2q}\eta-384H_0^{2}\Upsilon^{2q}z\eta\\
\nonumber &-& 288c_{2}H_0^{2}\Upsilon^{2q}\eta
-192H_0^{2}\Upsilon^{2q} z^{2}\eta-144 c_{2} H_0^{2} \Upsilon^{2q} z^{2}\eta\\
\nonumber &-& 16\sqrt{6}c_{3} \sqrt{H_0^{2}\Upsilon^{2 +2 q}}
\eta+\big(\sqrt{6H_0^{2}\Upsilon^{2+2 q}}(3 c_{2}-8 \eta)\\
\nonumber &+&2 c_{3}-2c_{1}\big)2 c_{3}\ln(6H_0^{2}\Upsilon^{2 + 2
q})
+2 c_{3}^{2}\ln(6H_0^{2}\Upsilon^{2+2q})^{2}\bigg)\bigg\} \\
\nonumber &\times& \bigg\{\bigg(3 \big(c_{3}-c_{1}+\sqrt{6} c_{2}
\sqrt{H_0^{2}\Upsilon^{2+2q}}\big)+c_{3}\ln[6 H_0^{2}\Upsilon^{2+2 q}]\bigg)\\
\nonumber &\times&\bigg(2c_{3}-2c_{1}+8 \sqrt{6}
\sqrt{H_0^{2}\Upsilon^{(2
+ 2 q)}}+3 \sqrt{6} c_{2}\sqrt{H_0^{2}\Upsilon^{2 + 2 q}} \\
\nonumber &+& 2 c_{3}\ln(6 H_0^{2}\Upsilon^{2 +
2q})\bigg)^{2}\bigg\}^{-1}-\bigg[\bigg\{\bigg
(2 \eta+\bigg(\Upsilon^{-2 (1+q)}\\
\nonumber &\times& \sqrt{H_0^{2}\Upsilon^{2 + 2 q}}\bigg(2 c_{3}-2
c_{1}+2c_{3}\ln(6H_0^{2}\Upsilon^{2+2 q})+ 3 \sqrt{6}c_{2} \\
\nonumber &\times&\sqrt{H_0^{2}\Upsilon^{2 + 2
q}}\bigg)\bigg)\bigg)\bigg(\sqrt{6}
H_0^{2}\bigg)^{-1}\bigg\}\bigg\{\bigg(2+\bigg(\Upsilon^{-2 (1 + q)}\\
\nonumber &\times&\sqrt{H_0^{2}\Upsilon^{2 + 2 q}} \bigg(-2c_{1}+2
c_{3} +3 \sqrt{6} c_{2}\sqrt{H_0^{2}\Upsilon^{2 + 2 q}} \\
\nonumber &+& 2 c_{3} \ln(6H_0^2 \Upsilon^{2 + 2
q})\bigg)\bigg)\bigg)\bigg(\sqrt{6}
H_0^{2}\bigg)^{-1}\bigg\}^{-1}\bigg]\bigg]\bigg]\bigg[2 c_{3}-2 c_{1}\\
\label{63}&+&3 \sqrt{6} c_{2}\sqrt{H_0^{2}\Upsilon^{2 + 2 q}}+2
c_{3} \ln[6 H_0^{2}\Upsilon^{2 + 2 q}]\bigg]^{-1}.
\end{eqnarray}
Figure \textbf{6} demonstrates a positively increasing trajectory of
$\nu_{s}^{2}$, implying the stability of GGDE model for all values
of $\eta$.
\begin{figure}[H]\center
\epsfig{file=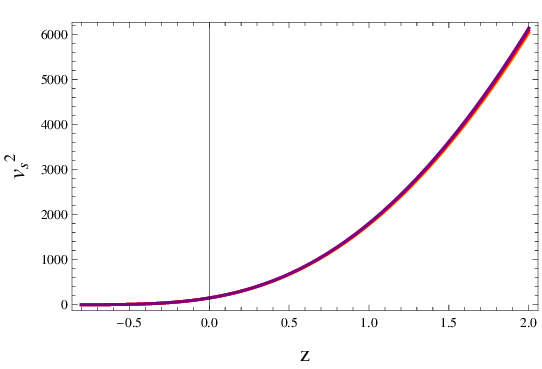,width=.6\linewidth}\caption{Graph of $\nu_s^{2}$
with $z$.}
\end{figure}

Stability analysis is used to investigate the conditions under which
cosmic formations maintain their stability in the presence of
various oscillation patterns. The squared speed of sound determines
the rate at which pressure waves travel through a medium. Its
signature plays a crucial role in discussing the stability of the
reconstructed GGDE model. Here, it is worth highlighting that we
have achieved increased stability behavior, thanks to the inclusion
of non-metricity, in contrast to both GR and other MT. It is also
noted that certain characteristics of the original GDE in $f(Q)$
cosmology differ from those in GR.

\section{Conclusions}

The phenomena of reconstruction in MT of gravity provides a helpful
instrument for creating a workable DE model that predicts the course
of cosmic progress. In this situation, we evaluate and contrast the
respective energy densities of DE and MT. This enables the
correspondence technique to lead to the required generic function of
underlying gravity theory. Many researchers have employed this
scheme to derive feasible DE models. In this paper, we have studied
the GGDE model in $f(Q)$ gravity. Firstly, we have reconstructed the
GGDE $f(Q)$ gravity model using the correspondence scheme. We have
used FRW model with power-law form of the scale factor for the
interacting case. We have then examined the evolution trajectories
of the EoS, the $(\omega_{D}-\omega^{\prime}_{D})$ and
$(r-s)$-planes. Finally, we have explored the solutions of the DE
model . The main results are given as follows.
\begin{itemize}
\item
The resulting GGDE $f(Q)$ gravity model demonstrates an increasing
trend for both $z$ as well as $Q$, signifying the realistic nature
of the reconstructed model (Figure \textbf{1}).
\item
The energy density exhibits positive behavior and increases, while
the pressure displays negative behavior. This behavior coincides
with the characteristics of DE (Figure \textbf{2}).
\item
We have found that the EoS characterizes the late universe involving
phantom field DE. We have also observed that the EoS takes more
negative values below $-1$ as the interaction parameter $\eta$
increases (Figure \textbf{3}). Our findings thus support the present
accelerated cosmic behavior.
\item
The evolutionary behavior of
($\omega_{D}$-$\omega^{\prime}_{D}$)-plane represents the freezing
region for all values of $\eta$ (Figure \textbf{4}). This confirms
that the interacting GGDE $f(Q)$ gravity model implies more
accelerated expanding universe.
\item
The $(r-s)$-plane represents Chaplygin gas model for different
values of $\eta$ (Figure \textbf{5}).
\item
We have found that the $\nu_{s}^{2}$ is positive and hence GGDE
$f(Q)$ gravity model is stable for all values of $\eta$ (Figure
\textbf{6}).
\end{itemize}

We have observed that the GGDE $f(Q)$ model demonstrates stable
characteristics and maintains consistent behavior with the current
accelerated expansion paradigm of the universe. The phantom-like
behavior of the cosmos is noted to anticipate a more accelerated
regime, which might ultimately lead to the big-rip scenario or the
universe's present accelerated condition. We noticed that the
results are consistent with the current observational data \cite{1a}
given as
\begin{eqnarray}\nonumber
\omega_{D}&=&-1.023^{+0.091}_{-0.096}\quad(\text{Planck
TT+LowP+ext}),\\\nonumber
\omega_{D}&=&-1.006^{+0.085}_{-0.091}\quad(\text{Planck
TT+LowP+lensing+ext}),\\\nonumber
\omega_{D}&=&-1.019^{+0.075}_{-0.080}\quad (\text{Planck TT, TE,
EE+LowP+ext}).
\end{eqnarray}
These values have been determined through the application of various
observational techniques at a 95\% confidence level. It is
noteworthy to mention here that our results coincide with the
findings from the reconstructed quantum chromodynamics ghost model
in $f(T)$ gravity \cite{1b} and the $f(R,T)$ theory \cite{1c}. We
would like to emphasize that our results coincide with the latest
theoretical observational data examined by Myrzakulov et al.
\cite{1d}.

\section*{Appendix A: Calculation of $Q=-Q_{\mu\gamma\psi}\mathbb{P}^{\mu\gamma\psi}$}
\renewcommand{\theequation}{A\arabic{equation}}
\setcounter{equation}{0}

We can write from Eq.\eqref{11} as
\begin{eqnarray}\nonumber
-g^{\gamma\psi}\mathbb{L}^{\mu}_{\;\nu\gamma}\mathbb{L}^{\nu}_{\;\psi\mu}
&=&-\frac{1}{4}g^{\gamma\psi}g^{\mu\lambda}g^{\nu\rho}(Q_{\mu\nu\lambda}
+Q_{\nu\lambda\mu}-Q_{\lambda\mu\nu})(Q_{\mu\psi\rho}+Q_{\psi\rho\mu}-Q_{\rho\mu\psi})\\
\nonumber&=&-\frac{1}{4}(Q^{\psi\rho\mu}+Q^{\rho\mu\psi}-Q^{\mu\psi\rho})
(Q_{\mu\psi\rho}+Q_{\psi\rho\mu}-Q_{\rho\mu\psi})\\\label{A6}
&=&-\frac{1}{4}(2Q^{\mu\psi\rho}Q_{\rho\mu\psi}
-Q^{\mu\psi\rho}Q_{\mu\psi\rho}), \\\nonumber
g^{\gamma\psi}\mathbb{L}^{\mu}_{\;\nu\mu}\mathbb{L}^{\nu}_{\;\gamma\psi}
&=&\frac{1}{4}g^{\gamma\psi}g^{\nu\rho}Q_{\nu}(Q_{\psi\gamma\rho}+Q_{\gamma\rho\psi}
-Q_{\rho\psi\gamma})\\\label{A7}
&=&\frac{1}{4}Q^{\rho}(2\tilde{Q}_{\rho}-Q_{\rho}).\label{A8}
\end{eqnarray}
Inserting these values in Eq.\eqref{22}, we obtain
\begin{equation}\label{A8}
Q=-\frac{1}{4}(-Q^{\mu\psi\rho}Q_{\mu\psi\rho}+2Q^{\mu\psi\rho}Q_{\rho\mu\psi}
-2Q^{\rho}\tilde{Q}_{\rho}-Q^{\rho}Q_{\rho}).
\end{equation}
We can also find $Q$ as follows. Using Eq.\eqref{11} in \eqref{25},
we have
\begin{eqnarray}\nonumber
\mathbb{P}^{\mu\gamma\psi}&=&\frac{1}{4}\big[-Q^{\mu\gamma\psi}+Q^{\gamma\mu\psi}
+Q^{\psi\mu\gamma}+Q^{\gamma\mu\psi}-\tilde{Q}_{\mu}g^{\gamma\psi}
+Q^{\mu}g^{\gamma\psi}\\\label{A9}
&-&\frac{1}{2}(Q^{\psi}g^{\mu\gamma}+Q^{\gamma}g^{\mu\psi})\big],\\\nonumber
-Q_{\mu\gamma\psi}\mathbb{P}^{\mu\gamma\psi}&=&-\frac{1}{4}\big[-Q_{\mu\gamma\psi}
Q^{\mu\gamma\psi}+Q_{\mu\gamma\psi}Q^{\psi\mu\gamma}+Q_{\mu\gamma\psi}
Q^{\gamma\mu\psi}-Q_{\mu\gamma\psi}\tilde{Q}_{\mu}g^{\gamma\psi}\\\nonumber
&+&Q_{\mu\gamma\psi}Q^{\mu}g^{\gamma\psi}-\frac{1}{2}Q_{\mu\gamma\psi}
(Q^{\psi}g^{\mu\gamma}+Q^{\gamma}g^{\mu\psi})\big]\\\nonumber
&=&-\frac{1}{4}\big[-Q_{\mu\gamma\psi}Q^{\mu\gamma\psi}
+2Q_{\mu\gamma\psi}Q^{\gamma\mu\psi}-2\tilde{Q}^{\mu}Q_{\mu}+Q^{\mu}Q_{\mu}\big]\\
\label{A10}&=&Q.
\end{eqnarray}
We have used the relations,
$Q_{\mu\gamma\psi}Q^{\gamma\mu\psi}=Q_{\mu\gamma\psi}Q^{\psi\mu\gamma}$,
to obtain the above result since
$Q_{\mu\gamma\psi}Q^{\gamma\mu\psi}=Q_{\mu\psi\gamma}Q^{\gamma\mu\psi}
=Q^{\mu\psi\gamma}Q_{\gamma\mu\psi}=Q^{\psi\gamma\mu}Q_{\mu\psi\gamma}
=Q_{\mu\gamma\psi}Q^{\psi\mu\gamma}$.

\section*{Appendix B: Evaluation of $\delta Q$}
\renewcommand{\theequation}{B\arabic{equation}}
\setcounter{equation}{0}

First we write down different formulae of $Q$ as follows
\begin{eqnarray*}\label{B1}
Q_{\mu\gamma\psi}&=&\nabla_{\mu}g_{\gamma\psi}, \\\label{B2}
Q^{\mu}_{~\gamma\psi}&=&g^{\mu\nu}Q_{\nu\gamma\psi}=g^{\mu\nu}\nabla_{\nu}g_{\gamma\psi}
=\nabla^{\mu}g_{\gamma\psi}, \\\label{B3}
Q^{~~\gamma}_{\mu~~\psi}&=&g^{\gamma\rho}Q_{\mu\rho\psi}=g^{\gamma\rho}\nabla_{\mu}
g_{\rho\psi}=-g_{\rho\psi}\nabla_{\mu}g^{\gamma\rho}, \\\label{B4}
Q_{\mu\gamma}^{\quad\psi}&=&g^{\psi\rho}Q_{\mu\gamma\rho}=g^{\psi\rho}\nabla_{\mu}
g_{\gamma\rho}=-g_{\gamma\rho}\nabla_{\mu}g^{\psi\rho},\\\label{B5}
Q^{\mu\gamma}_{\quad\psi}&=&g^{\mu\nu}g^{\gamma\rho}\nabla_{\nu}g_{\rho\psi}
=g^{\gamma\rho}\nabla^{\mu}g_{\rho\psi}=-g_{\rho\psi}\nabla^{\mu}g^{\gamma\rho},
\\\label{B6}
Q^{\mu~~\psi}_{~~\gamma}&=&g^{\mu\nu}g^{\psi\rho}\nabla_{\nu}g_{\gamma\rho}
=g^{\psi\rho}\nabla^{\mu}g_{\gamma\rho}
=-g_{\gamma\rho}\nabla^{\mu}g^{\psi\rho}, \\\label{B7}
Q_{\mu}^{\quad\gamma\psi}&=&g^{\gamma\rho}g^{\psi\sigma}\nabla_{\mu}g_{\rho\sigma}
=-g^{\gamma\rho}g_{\rho\sigma}\nabla_{\mu}g^{\psi\sigma}
=-\nabla_{\mu}g^{\gamma\psi}, \\\label{B8}
Q^{\mu\gamma\psi}&=&\nabla^{\mu}g^{\gamma\psi}.
\end{eqnarray*}
By using Eq.\eqref{A8}, we find the variation of $Q$ as
\begin{eqnarray}\nonumber
\delta
Q&=&-\frac{1}{4}\delta(-Q^{\mu\psi\rho}Q_{\mu\psi\rho}
+2Q^{\mu\psi\rho}Q_{\rho\mu\psi}-2Q^{\rho}\tilde{Q}_{\rho}+Q^{\rho}Q_{\rho})
\\\nonumber &=&-\frac{1}{4}(-\delta
Q^{\mu\psi\rho}Q_{\mu\psi\rho}-Q^{\mu\psi\rho}\delta
Q_{\mu\psi\rho}+2\delta Q^{\mu\psi\rho}Q_{\rho\mu\psi}
\\\nonumber &+&2\delta Q^{\mu\psi\rho}\delta Q_{\rho\mu\psi}-2\delta
Q^{\rho}\tilde{Q}_{\rho}-2Q^{\rho}\delta\tilde{Q}_{\rho}+\delta
Q^{\rho}Q_{\rho}+Q^{\rho}\delta Q_{\rho})
\\\nonumber &=&-\frac{1}{4}\bigg[Q_{\mu\psi\rho}\nabla^{\mu}\delta
g^{\psi\rho}-Q^{\mu\psi\rho}\nabla_{\mu}\delta
g_{\psi\rho}-2Q_{\rho\mu\psi}\nabla^{\mu}\delta g^{\psi\rho}
\\\nonumber &+&2Q^{\mu\psi\rho}\nabla_{\rho}\delta
g_{\mu\psi}-2\tilde{Q}_{\rho}\delta(-g_{\gamma\psi}\nabla^{\rho}g^{\gamma\psi})-2Q^{\rho}\delta(\nabla^{\lambda}g_{\rho\lambda})
\\\nonumber
&+&Q_{\rho}\delta(-g_{\gamma\psi}\nabla^{\rho}g^{\gamma\psi})+Q^{\rho}\delta(-g_{\gamma\psi}\nabla_{\rho}g^{\gamma\psi})\bigg]
\\\nonumber &=&-\frac{1}{4}\bigg[Q_{\mu\psi\rho}\nabla^{\mu}\delta
g^{\psi\rho}-Q^{\mu\psi\rho}\nabla_{\mu}\delta
g_{\psi\rho}-2Q_{\rho\mu\psi}\nabla^{\mu}\delta g^{\psi\rho}
\\\nonumber &+&2Q^{\gamma\psi\rho}\nabla_{\rho}\delta
g_{\gamma\psi}+2\tilde{Q}_{\rho}\nabla^{\rho}g^{\gamma\psi}\delta
g_{\gamma\psi}+2\tilde{Q}_{\rho}g_{\gamma\psi}\nabla^{\rho}\delta
g^{\gamma\psi}
\\\nonumber &-&2Q^{\rho}\nabla^{\lambda}\delta
g_{\rho\lambda}-Q_{\rho}\nabla^{\rho}g^{\gamma\psi}\delta
g_{\gamma\psi}-Q_{\rho}g_{\gamma\psi}\nabla^{\rho}\delta
g^{\gamma\psi}
\\\nonumber &-&Q^{\rho}\nabla_{\rho}g^{\gamma\psi}\delta
g_{\gamma\psi}-Q^{\rho} g_{\gamma\psi}\nabla_{\rho}\delta
g^{\gamma\psi}\bigg]
\\\nonumber &=&-\frac{1}{4}\bigg[Q_{\mu\psi\rho}\nabla^{\mu}\delta
g^{\psi\rho}-Q^{\mu\psi\rho}\nabla_{\mu}\delta
g_{\psi\rho}-2Q_{\rho\mu\psi}\nabla^{\mu}\delta g^{\psi\rho}
\\\nonumber &+&2Q^{\mu\psi\rho}\nabla_{\rho}\delta
g_{\mu\psi}+2\tilde{Q}_{\rho}\nabla^{\rho}g^{\gamma\psi}\delta
g_{\gamma\psi}+2\tilde{Q}_{\rho}g_{\gamma\psi}\nabla^{\rho}\delta
g^{\gamma\psi}
\\\nonumber &-&2Q^{\rho}\nabla^{\lambda}\delta
g_{\rho\lambda}-Q_{\rho}\nabla^{\rho}g^{\gamma\psi}\delta
g_{\gamma\psi}-Q_{\rho}g_{\gamma\psi}\nabla^{\rho}\delta
g^{\gamma\psi}
\\\label{B9}
&-&Q^{\rho}\nabla_{\rho}g^{\gamma\psi}\delta g_{\gamma\psi}-Q^{\rho}
g_{\gamma\psi}\nabla_{\rho}\delta g^{\gamma\psi}\bigg].
\end{eqnarray}
Since $\delta g_{\gamma\psi}=- g_{\gamma\mu}\delta
g^{\mu\beta}g_{\beta\psi}$, thus we can write
\begin{eqnarray*}\label{B10}
-Q^{\mu\psi\rho}\nabla_{\mu}\delta g_{\psi\rho}
&=&-Q^{\mu\psi\rho}\nabla_{\mu}(-g_{\psi\lambda}\delta
g^{\lambda\theta}g_{\theta\rho})=2Q^{\mu\psi}_{\quad
\theta}Q_{\mu\psi\lambda}\delta
g^{\lambda\theta}+Q_{\mu\lambda\theta}\nabla^{\alpha}g^{\lambda\theta}
\\\label{B11}
&=&2Q^{\mu\sigma}_{\quad\psi}Q_{\mu\sigma\gamma}\delta
g^{\gamma\psi}+Q_{\mu\psi\rho}\nabla^{\alpha}g^{\psi\rho},
\\\label{B12} 2Q^{\mu\psi\rho}\nabla_{\rho}\delta
g_{\mu\psi}&=&-4Q_{\gamma}^{\quad\sigma\rho}Q_{\rho\sigma\psi}\delta
g^{\gamma\psi}-2Q_{\psi\rho\alpha}\nabla^{\mu}\delta g^{\psi\rho},
\\\label{B13}
-2Q^{\rho}\nabla^{\lambda}\delta
g_{\rho\lambda}&=&2Q^{\mu}Q_{\psi\mu\gamma}\delta
g^{\gamma\psi}+2Q_{\gamma}\hat{Q}_{\psi}\delta g^{\gamma\psi}
+2Q_{\psi}g_{\mu\rho}\nabla^{\mu}g^{\psi\rho}.
\end{eqnarray*}
Consequently, Eq.\eqref{B9} takes the form
\begin{eqnarray}\nonumber
\delta Q&=&-\frac{1}{4}\bigg[Q_{\mu\psi\rho}\nabla^{\mu}\delta
g^{\psi\rho}+2Q^{\mu\sigma}_{\quad\nu}Q_{\mu\sigma\gamma}\delta
g^{\gamma\psi}+Q_{\mu\psi\rho}\nabla^{\mu}\delta
g^{\psi\rho}-2Q_{\rho\mu\psi}\nabla^{\mu}\delta
g^{\psi\rho}\\\nonumber
&-&4Q_{\gamma}^{\quad\sigma\rho}Q_{\rho\sigma\psi}\delta
g^{\gamma\psi}-2Q_{\psi\rho\mu}\nabla^{\mu}\delta
g^{\psi\rho}+2\hat{Q}^{\rho}Q_{\rho\gamma\psi}\delta
g^{\gamma\psi}+2\hat{Q}_{\mu}g_{\psi\rho}\nabla^{\mu}\delta
g^{\psi\rho}
\\\nonumber
&+&2Q^{\psi\mu\gamma}\delta
g^{\gamma\psi}+2Q_{\gamma}\hat{Q}_{\psi}\delta
g^{\gamma\psi}+2Q_{\nu}g_{\mu\rho}\nabla^{\alpha}g^{\psi\rho}
-Q^{\rho}Q_{\rho\gamma\psi}\delta g^{\gamma\psi}\\\nonumber
&-&Q_{\mu}g_{\psi\rho}\nabla^{\mu}g^{\psi\rho}-Q^{\rho}Q_{\rho\gamma\psi}\delta
g^{\gamma\psi}-Q_{\mu}g_{\psi\rho}\nabla^{\mu}g^{\psi\rho}\bigg].
\end{eqnarray}
Using the following relations in the above equation
\begin{equation*}
2P_{\mu\psi\rho}=-\frac{1}{4}\bigg[2Q_{\mu\psi\rho}-2Q_{\rho\mu\psi}-2Q_{\psi\rho\mu}
+2(\hat Q_{\mu}- Q_{\mu})g_{\psi\rho}+2Q_{\psi}g_{\mu\rho}\bigg],
\end{equation*}
\begin{eqnarray*}\nonumber
4(P_{\gamma\mu\nu}Q_{\psi}^{~~\mu\nu}-2Q^{\mu\nu}_{\quad\gamma}P_{\mu\nu\psi})&=&2Q^{\mu\nu}_{\quad\psi}
Q_{\mu\nu\psi}-4Q_{\gamma}^{~~\mu\nu}Q_{\nu\mu\psi}+2\tilde
Q^{\mu}Q_{\mu\gamma\psi}\\\label{B17}
&+&2Q^{\mu}Q^{\psi\mu\gamma}+2Q_{\gamma}\tilde
Q_{\psi}-Q^{\alpha}Q_{\mu\gamma\psi},
\end{eqnarray*}
we obtain
\begin{eqnarray}\label{B15}
\delta
Q&=&2P_{\mu\psi\rho}\nabla^{\mu}g^{\psi\rho}-(P_{\gamma\mu\nu}Q_{\psi}^{\quad\mu\nu}
-2Q^{\mu\nu}_{\quad\gamma}P_{\mu\nu\psi})\delta g^{\gamma\psi}.
\end{eqnarray}

\section*{Appendix C: \textbf{Computation} of $Q=6H^{2}$}
\renewcommand{\theequation}{C\arabic{equation}}
\setcounter{equation}{0}

Using Eq.\eqref{A10}, we obtain
\begin{equation}\label{C1}
Q=-\frac{1}{4}\big[-Q_{\mu\gamma\psi}Q^{\mu\gamma\psi}
+2Q_{\mu\gamma\psi}Q^{\gamma\mu\psi}-2\tilde{Q}^{\mu}Q_{\mu}+Q^{\mu}Q_{\mu}\big].
\end{equation}
In Appendix \textbf{B}, we have used the relations
\begin{eqnarray}\nonumber
-Q_{\mu\gamma\psi}Q^{\mu\gamma\psi}&=&\nabla_{\mu}g_{\gamma\psi}\nabla^{\mu}
g^{\gamma\psi}=12H^{2}, \\\nonumber
Q_{\mu\gamma\psi}Q^{\gamma\mu\psi}&=&-\nabla_{\mu}g_{\gamma\psi}\nabla^{\gamma}
g^{\mu\psi}=0, \\\nonumber
\tilde{Q}^{\mu}Q_{\mu}&=&(g_{\gamma\rho}\nabla_{\mu}g^{\gamma\rho})(\nabla_{\nu}
g^{\mu\nu})=0, \\\nonumber
Q^{\mu}Q_{\mu}&=&(g_{\rho\gamma}\nabla_{\mu}g^{\rho\gamma})(g_{\sigma\psi}\nabla^{\mu}
g^{\sigma\psi})=-36H^{2}.
\end{eqnarray}
Inserting these values in \eqref{C1}, we have
\begin{equation}\nonumber
Q=6H^{2}.
\end{equation}\\
\textbf{Data Availability:} No data was used for the research
described in this paper.

\end{document}